\documentclass[12pt]{article}
\usepackage{amsmath}
\usepackage{amsfonts}
\usepackage{amsthm}
\def\acts{\triangleright}
\def\Z{{\mathbb Z}}
\def\C{{\mathbb C}}
\def\R{{\mathbb R}}
\def\CA{{\mathcal A}}

\def\CH{{\mathcal H}}

\def\CU{{\mathcal U}}
\DeclareMathSymbol\crossrt{\mathrel}{AMSb}{"6E}
\DeclareMathSymbol\crosslt{\mathrel}{AMSb}{"6F}

\def\id{{\mathrm i}{\mathrm d}}
\def\ts{\otimes}
\newcommand{\kett}[1]{|#1\rangle\!\rangle} %% ket 2-vector

\def\oh{\frac{1}{2}}
\def\half{\oh}

\def\acts{\triangleright}

\def\ts{\otimes}
\def\ket#1{| #1 \rangle}

\def\nn{\nonumber}

\newcommand{\up}{{\mathord{\uparrow}}} %% `up' spinors
\newcommand{\dn}{{\mathord{\downarrow}}} %% `down' spinors
\newtheorem{lemma}{Lemma}[section]
\newtheorem{theorem}[lemma]{Theorem}
\newtheorem{proposition}[lemma]{Proposition}

\newtheorem{definition}[lemma]{Definition}

\newtheorem{remark}[lemma]{Remark}

\title{Equivariant Lorentzian Spectral Triples}
\author{ Mario Paschke \\
{\em Max-Planck-Institut f\"ur Mathematik in den Naturwissenschaften} \\
{\em Inselstr.\ 22, 04103 Leipzig, Germany} \\
\and Andrzej Sitarz\thanks{Alexander von Humboldt Fellow}
\thanks{Partially supported by MNII Grant 115/E-343/SPB/6.PR UE/DIE 50/2005--2008}\\
{\em Institute of Physics, Jagiellonian University}\\
{\em Reymonta 4, 30-059 Krak\'ow, Poland} \\
and \\
{\em Mathematisches Institut, Heinrich-Heine-Universit\"at}\\
{\em Universit\"atsstrasse 1, 40225 D\"usseldorf, Germany}
}
\begin{document}
\maketitle
\begin{abstract}
\noindent
We present examples of equivariant noncommutative Lorentzian
spectral geometries. The equivariance with respect to a compact
isometry group (or quantum group) allows to construct the
algebraic data of a version of spectral triple geometry
adapted to the situation of an indefinite metric. The
spectrum of the equivariant Dirac operator is calculated.
\end{abstract}


\section{Introduction}
Equivariance under compact quantum groups has proven to be a very efficient
tool to construct new explicit examples of spectral triples. The equivariance
of the spectral triple algebra representation fixes the Hilbert space as
a representation space of the quantum group and enables to diagonalize the
Dirac operator on finite-dimensional subspaces invariant under its action.
The eigenvalues of the Dirac can be, in turn, computed with the help
of the order-one-condition.

In contrast to that situation, Lorentzian spectral triples, and in particular
classical Lorentzian spin manifolds, generically admit {\em noncompact} isometry
groups, with the relevant representations -- occurring as eigenspaces of the
Lorentzian Dirac operator --  being infinite dimensional. From the point of view
of a physicist this large space of solutions of the Dirac-equation (to a given
mass) is of course desired. Moreover, as has been pointed out in \cite{KP}, one
may reconstruct the complete information about the metric and the spin structure
of a (commutative) Lorentzian spin manifold from the space of solution to the
Dirac-equation to a fixed mass.  In that work it was also shown that certain
Lorentzian spectral triples can, in principle, be constructed explicitly by
exploiting their isometries.

However, the procedure adopted there appears much less systematic than the
corresponding one for the Euclidean spectral triples. For instance, the
decomposition of the Hilbert space into irreducible representations of the
isometry group ($SL(2,\mathbb{R})$ in that case) is not derived systematically
but rather put in by hand, in order to circumvent some technical problems
related to the infinite dimension of the relevant representations.  Moreover,
unlike in the case of compact isometry groups, not all series of representations
appear in the decomposition of the Hilbert space.

Equivariant Lorentzian spectral triples are very interesting in a physical
context for the model-building, especially since there are many
known deformations of the Lorentz and Poincar\'e groups. From the
mathematical point of view, equivariance is still the most efficient
technical tool to construct the infinite-dimensional eigenspaces of
the Dirac operator also in this situation. It appears, that this tool
seems indispensable for the construction of genuine
{\em noncommutative} examples of Lorentzian spectral triples.

As pointed out above, the main problem when dealing with equivariance
in the Lorentzian case is the noncompactness of the full isometry groups.
However, considering the few examples of compact Lorentzian manifolds,
which admit compact isometry groups, we can observe that these groups
are necessarily much smaller then the isometry group of the same manifold
when equipped with the (maximally symmetric) Riemannian metric.

As the construction of (Riemannian) spectral triples via equivariance uses
the full isometry group, it may then seem at first sight that the (reduced)
isometry groups for the Lorentzian case may not suffice for a systematic
construction of a Lorentzian spectral triple.

In this paper we shall demonstrate that this is not the case, as
the data of Lorentzian spectral triples involve an additional
operator, the fundamental symmetry $\beta$. The equivariance condition
for $\beta$ together with that for $J$,$D$ and the representation of $\CA$
indeed does provide enough equations to systematically construct a (compact)
Lorentzian spectral triple (up to a scale) with classical, compact
isometry groups.

The problem of constructing equivariant Lorentzian compact
spectral triples can be in some cases reduced to the problem
of finding a Euclidean spectral geometry but with an isometry
group smaller than the maximally allowed. We shall see, that
in some cases, like, for instance, in the $SU_q(2)$ case,
it is still not clear whether the full spectral data (including
the reality structure) can be obtained.

\section{Axioms for Lorentzian spectral triples}

We begin with the  axioms for real Lorentzian
spectral geometries, which shall be discussed thoroughly in \cite{PaReVe}.
They are analogous to those for the Euclidean
case \cite{CoRe} and, in fact, based on the idea that to each
Lorentzian spectral geometry, it should be possible to
associate a corresponding Euclidean one and thereby a well-defined index map
\cite{PaRe}.

\begin{definition}
{\em A geometric real (odd or even) Lorentzian spectral
triple} of signature $(1,q)$ is given by the data
$(\CA,\pi,\CH,D,J,\gamma,\beta)$, where:
\begin{itemize}
\item $\CA$ is an involutive algebra, $\pi$ its faithful bounded star
      representation on a Hilbert space $\CH$,
\item in the even case, $1+q \in 2\Z$,
   $\gamma=\gamma^\dagger$, $\gamma^2=1$ is a $\Z_2$ grading,
   commuting with the representation of $\CA$,
\item $J$ is an antilinear isometry such that:
\begin{equation}
[ J\pi(a)J^{-1}, \pi(b)] = 0, \;\;\; \forall a,b \in \CA,
\end{equation}
\item $\beta=-\beta^\dagger$, $\beta^2=-1$ is the
$\Z_2$-grading associated with the Krein-space structure,
commuting with the representation of the algebra $\CA$,
\item $D$ an unbounded, densely defined operator, which
is $\beta$-selfadjoint, that is: $D^\dagger = \beta D \beta$,
and such that $[D,\pi(a)]$ is bounded for every $a \in \CA$,
and $D\gamma = -\gamma D$.
\item The operator
\begin{equation}
\langle D \rangle = \sqrt{\oh(DD^\dagger+D^\dagger D)}\label{laplace}
\end{equation}
has compact resolvent. In addition it is required that
$[\langle D \rangle ,[D,\pi(a)]]$ is bounded for all $a\in\CA$.
\item The grading, reality structure and the Dirac operator satisfy:
\begin{equation}
D J = \epsilon JD, \;\;\; J^2=\epsilon',
\;\;\;  J \gamma = \epsilon'' \gamma J.
\end{equation}
where $\epsilon, \epsilon', \epsilon''$ are $\pm1$ depending on
$1-q$ modulo 8 according to the following rules:

\begin{center}
\begin{tabular}{|c|c|c|c|c|c|c|c|c|}
\hline
$1-q$ mod 8  & 0 & 1 & 2  & 3  & 4  & 5  & 6  & 7 \\ \hline
$\epsilon$   & + & + & +  & -- & +  & +  & +  & -- \\ \hline
$\epsilon'$  & + & + & +  & -- & -- & -- & -- & +  \\ \hline
$\epsilon''$ & + &   & -- &    & +  &    & -- &    \\ \hline
\end{tabular}
\end{center}

\item the Krein-space structure satisfies:
\begin{equation}
%%\beta \gamma = (-1)^p \gamma \beta, \;\;\;\;
%%\beta J = - (-1)^{\oh p(p-1)} \epsilon^p J \beta,
\beta \gamma = - \gamma \beta, \;\;\;\;
\beta J = -  \epsilon^p J \beta,
\end{equation}

\item The Dirac operator satisfies the order-one condition:
\begin{equation}
\left[ J\pi(a)J^{-1}, [D,\pi(b)] \right] = 0, \;\;\; \forall a,b
\in \CA.
\label{orderone}
\end{equation}
\item There exists a Hochschild cycle of dimension $n=1+q$,
valued in $\CA^o \otimes \CA$,
$$ c = a_{i_0}^o \otimes a_{i_o} \otimes a_{i_1}\otimes \cdots \otimes a_{i_n}, $$
such that:
$$ J\pi(a_{i_0})J^{-1} \pi(a_{i_1}) [D, \pi(a_{i_1})] \ldots [D, \pi(a_{i_n})]
= \begin{cases} \gamma & n \;\; \hbox{even} \\ 1 & n \;\; \hbox{odd}
\end{cases}, $$
\item  We say that the Lorentzian spectral triple has
time-orientation if there exist $a_i^o, a_i,b_i \in \CA$
such that
\begin{equation}
\beta = \sum \limits_i J\pi(a_i^o)J^{-1} \pi(a_i) [D,\pi(b_i)].
\label{timeorient}
\end{equation}
\end{itemize}
If we do not assume existence of $J$, we have a spectral triple
without real structure.\footnote{Note that in that case it is
difficult to say whether we are in a typically Lorentzian case,
with one "time-like" noncommutative direction. We might as
well be in the case with a signature $(p,q)$, $p-q=1 \mod 2$.}
\end{definition}

We restrict ourselves only to the algebraic requirement and we
refer the reader to the papers \cite{PaReVe,PaRe} for details
on further analytic requirements like summability, finiteness
conditions as well as the Poincar\'e duality, which can be quite
similarly postulated here, as it is done in the Euclidean case
(see \cite{BFV}, for instance).

\begin{remark}
The definition can be in a straightforward way extended to the
case of arbitrary signature $(p,q)$. Our notation is that the
Euclidean spectral geometry of dimension $0$ is identified with
the $(0,n)$ signature.
\end{remark}

The sign relations for the arbitrary signature have been studied
in the context of {\em spectral geometry axioms} by various
people \cite{GB, Meyer, Holfter}.

The basic motivation for the definition and the postulates for the
algebraic formulation comes from the differential geometry of
Lorentzian spin manifolds \cite{Baum}. The Lorentzian version of
the relations between the classical geometries and spectral
geometries for appropriate commutative algebras is discussed
elsewhere \cite{PaReVe}, we can quote:

\begin{lemma}(see \cite{PaReVe} for details)
Let $M$ be a compact Lorentzian spin manifold. Then taking
$\CA =C^\infty(M)$, $\CH$ to be the summable sections
of the spinor bundle, and $D$ the Dirac operator, $(\CA,\CH,D)$
is a Lorentzian spectral triple.

The operator $\beta$ could be identified with a specific choice of
a fundamental symmetry (denoted $J$ in \cite{Baum}). Different, but
normalized choices of $\beta$ lead to  equivalent Lorentzian spectral
triples in the sense that for the choices $\beta_1,\beta_2$
with $\beta_i^2 = -1$ for $i=1,2$, there exists a unitary
operator
$U\,\,: \CH_1 \mapsto \CH_2$ such that:

$$X_2 = U X_1 U^*,  \;\;\;\;
X\in \{\beta, J, \gamma, \pi(\CA) \},
$$
and
$$
[D_2, U a U^*] = U [D_1, \pi(a)]U^* \;\;\;\; \forall a \CA.
$$

\end{lemma}

In this paper we discuss the genuine noncommutative examples.

\subsection{Equivariance of Lorentzian triples}

We recall the notion of equivariance for spectral triples
\cite{Paschke-Th,Sitarz-eq}, extending it, in a natural way, to the
Lorentzian case.

\begin{definition}
Let $H$ be a Hopf algebra, acting on the algebra $\CA$.
A Lorentzian spectral triple is $H$-equivariant if:
\begin{itemize}
\item $H$ has a representation $\rho$ on a dense domain of $\CH$,
such that the representation of $\CA$ is equivariant:
$$ \rho(h) \pi(a) = \pi( h_{(1)} \acts a) \rho(h_{(2)}, $$
for all $a \in \CA, h\in H$, with the equality valid on a dense
domain of $\CH$,
\item $\gamma$, $\beta$ and $D$ commute with the representation $\rho$
of $H$
\item the reality structure $J$ is equivariant:
$$ J(\rho(Sh)^\dagger) J^{-1} = \rho(h), \;\; h \in H. $$
\end{itemize}
\end{definition}

\section{Lorentzian Spectral Triple for the Noncommutative Torus}

The Euclidean spectral triple of the noncommutative torus one
of the best known examples. Also in the Lorentzian case it
has been studied, albeit without the real structure
in \cite{Stroh}.

We shall investigate here it once again following exactly the
procedure of equivariance \cite{Sitarz-eq}, however, taking the
Lorentzian axioms for $(1,1)$ spectral geometry.

\subsection{Noncommutative Torus and its symmetries}

We recall the basic definitions,

\begin{definition}
Consider the Hilbert space $l^2(\Z^2)$ with the orthonormal basis
$\{ |n,m \rangle$, $n,m \in \Z \}$ and the unitary operators:
\begin{eqnarray*}
\pi(U) |n,m\rangle & = & |n+1,m\rangle,\nonumber \\
\pi(V) |n,m\rangle & = & \lambda^{-n} |n,m+1\rangle,
\label{rep}
\end{eqnarray*}
where $\lambda$ is complex number $|\lambda|=1$. The algebra
generated by these operators we shall call the algebra of functions on the
noncommutative torus.
\end{definition}
Note that we defined so far the algebra of polynomials
and we might complete it either to a Fr\'echet algebra or a
$C^*$ algebra.
\begin{proposition}
\label{remu2}
Let ${\bf u(1)\oplus u(1)}$ be  the
Lie algebra generated by two derivations on the noncommutative
torus:
\begin{eqnarray*}
\delta_1 \acts U = U, & \qquad \qquad & \delta_2 \acts U = 0, \\
\delta_1 \acts V = 0, & \qquad \qquad & \delta_2 \acts V = V.
\end{eqnarray*}
Then, with the representation:
\begin{eqnarray*}
\rho(\delta_1) | n,m \rangle  & = & n  | n,m \rangle, \\
\rho(\delta_2) | n,m \rangle  & = & m | n,m \rangle.
\end{eqnarray*}
we have the representation of the cross-product algebra of the functions
on the noncommutative torus by the symmetry algebra. (The latter being
the universal enveloping algebra of  ${\bf u(1)\oplus u(1)}$.)  Here, we
take as the dense subspace $V$ the linear space spanned by the
basis $| n,m \rangle$, $n,m \in \Z$.
\end{proposition}

To construct the real Lorentzian spectral triple we need
a grading $\gamma$ (which just doubles the Hilbert space)
the antilinear isometry $J$, and the Krein-space structure
$\beta$.

Note that for $(1,1)$ we have $\beta \gamma = - \gamma \beta$,
so by choosing the Hilbert space to be $\CH \otimes \C^2$ with
the diagonal representation $\hbox{diag}(\pi)$ and $\gamma$
diagonal with $\pm 1$:
$$ \gamma = \left( \begin{array}{cc} 1 & 0 \\ 0 & -1 \end{array} \right).$$
we have
$$ \beta = \left( \begin{array}{cc} 0 & 1 \\ -1 & 0   \end{array} \right).$$

Next, using the modular operator from the Tomita-Takesaki theory,
$$ J_0 |n,m\rangle  =  \lambda^{-nm}|-n,-m\rangle, $$ we obtain $J$
\footnote{Thus, we restrict ourselves to one choice of the spin
structure on the noncommutative torus, for details see
\cite{PaSiTo}},
by tensoring $J_0$ with a suitable matrix from $M_2(\C)$.
To satisfy the algebraic requirements of the real $(1,1)$ spectral
triple, we need to have:
$$ J^2=1, \;\;\;\; J \gamma=\gamma J, \;\;\;\; J\beta = -\beta J.$$
so it is clear that $J = J_0 \otimes \gamma$:
Taking $\gamma$ to be block diagonal we have:
\begin{equation}
J =  \left( \begin{array}{cc}
J_0 & 0 \\ 0 & - J_0
\end{array} \right).
\end{equation}

Next, let us come to the point of constructing the equivariant Dirac
operator, here again we repeat the steps from the Euclidean case.

Since $D$ anticommutes with $\gamma$, it must be of the form
$$ D =  \left( \begin{array}{cc}
0 & \partial_- \\ \partial_+ & 0
\end{array} \right).
$$

Taking into account that $D^\dagger =\beta D \beta $ is
selfadjoint we have:
\begin{equation}
(\partial_\pm)^\dagger =  \partial_\pm.
\label{Dirac-self}
\end{equation}

\begin{proposition}
\label{nct-dirac}
Every Dirac operator $D$, which is ${\bf u(1)}\oplus {\bf u(1)}$-equivariant
must be of the form given above, with $\partial_\pm$:
$$ \partial_\pm |n,m,\pm \rangle = d_{n,m}^\pm  |n,m,\mp \rangle, \;\;
n,m \in \Z. $$
\end{proposition}

This follows directly from the requirement $[D, \delta_i]=0$, $i=1,2$.
As we shall see, this assumption, together with other algebraic
requirements fixes $\partial$ up to a normalization factor.

\begin{lemma}
\label{torlem}
Any Dirac operator $D$, which has ${\bf u(1)}\oplus {\bf u(1)}$ as an isometry
and which is order-one (see (\ref{orderone})) on the $(1,1)$ spectral
geometry of the noncommutative torus, is defined by the set
of real coefficients $d_{n,m}^\pm$:
\begin{equation}
d_{n,m}^\pm = \tau_1^\pm n +  \tau_2^\pm m + \epsilon,
\end{equation}
\end{lemma}

\begin{proof}
First of all, using $JD=DJ$ we immediately get that the coefficients
$d^\pm_{m,n}$ must satisfy:
\begin{equation}
(d^\pm_{m,n})^* = - d^\pm_{-m,-n}.
\label{dirac-lo1}
\end{equation}
Then, from order-one condition we get:

\begin{align}
d_{n+1,m}^\pm & =   2 d_{n,m}^\pm-d_{n-1,m}^\pm, \\
d_{n+1,m}^\pm - d_{n,m}^\pm & =  d_{n+1,m-1}^\pm-d_{n,m-1}^\pm, \\
d_{n,m+1}^\pm - d_{n,m}^\pm & =   d_{n-1,m+1}^\pm-d_{n-1,m}^\pm, \\
d_{n,m+1}^\pm & =   2 d_{n,m}^\pm-d_{n,m-1}^\pm
\end{align}

The above recursion relations have solutions:
\begin{equation}
d_{n,m}^\pm = \tau_1^\pm n + \tau_2^\pm m + \epsilon^\pm,
\end{equation}
for arbitrary constants $\tau_i,\epsilon$.
Using (\ref{dirac-lo1}) we first see that:
$$(\tau_i^\pm)^* =  \tau_i^\pm, \; i=1,2,\;\;\;\;
(\epsilon^\pm)^* = -\epsilon^\pm.$$
On the other hand, using (\ref{Dirac-self}) we obtain:
$$(\tau_i^\pm)^* =  \tau_i^\pm,\; i=1,2,\;\;\;\;
(\epsilon^\pm)^* = \epsilon^\pm,$$
Therefore all $\tau_i^\pm$ must be real and $\epsilon^\pm=0$.
\end{proof}

We can now compute the operator $\langle D \rangle$ defined in
(\ref{laplace}), which comes out as
$$ \langle D \rangle =
\frac{1}{\sqrt{2}}
\left( \begin{array}{cc}
\sqrt{\partial_+^2 + \partial_-^2} & 0 \\
0 &    \sqrt{\partial_+^2 + \partial_-^2}
\end{array} \right).
$$
%%%%%%%%%%%%%%%%%%%%%%%%%%%%%%%%%%%%%%%%%%%%%%%%%%%%%%%%%%%%
%%%%%%%%%%%%%%%%%%%%%%%%%%%%%%%%%%%%%%%%%%%%%%%%%%%%%%%%%%%%
We shall have to investigate next whether this operator
has compact resolvent. Since  $\langle D \rangle$ is
already diagonalized,
$$
\langle D \rangle
|n,m,\pm \rangle =\frac{1}{\sqrt{2}}
\sqrt{(d_{n,m}^+)^2 +(d_{n,m}^-)^2}\, |n,m,\pm \rangle ,
$$

\begin{lemma}
\label{ellipticity}
The operator $\langle D \rangle$ has compact resolvent whenever
$$\tau^+_1\tau^-_2 \neq \tau^+_2\tau^-_1 .$$
\end{lemma}

\begin{proof}
We can rewrite the eigenvalues of $\langle D \rangle^2$ as
a quadratic form restricted to the $\Z^2$ lattice of vectors
in $\R^2$. On the basis vectors the form is:
$$
\left( \begin{array}{ll} (\tau_1^+)^2 + (\tau_1^-)^2 &
\tau_1^+ \tau_2^+ + \tau_1^- \tau_2^- \\
\tau_1^+ \tau_2^+ + \tau_1^- \tau_2^- &
 (\tau_2^+)^2 + (\tau_2^-)^2 \end{array} \right)
 $$
The form certainly non-negative but it is strictly
positive on non-zero vectors if and only if its
determinant is positive. This leads to:
$$ (\tau_1^+ \tau_2^- - \tau_1^- \tau_2^+)^2 > 0, $$
hence the above condition.

If $\tau_1^+ \tau_2^- = \tau_1^- \tau_2^+$, the eigenvalues
of $\langle D \rangle^2$ are:
$$ \left( (\tau_1^+)^2 + (\tau_1^-)^2 \right)
\left( n + \frac{\tau_1^+}{\tau_2^+} m \right)^2. $$
and it is clear that for any $\epsilon >0$ we can find
infinitely many pairs $(m,n)$ such that
$n + \frac{\tau_1^+}{\tau_2^+} m < \epsilon$.

Therefore for $\tau_1^+ \tau_2^- = \tau_1^- \tau_2^+$ the operator
$\langle D \rangle$ does not have a compact resolvent.
\end{proof}
%%%%%%%%%%%%%%%%%%%%%%%%%%%%%%%%%%%%%%%%%%%%%%%%
%%%%%%%%%%%%%%%%%%%%%%%%%%%%%%%%%%%%%%%%%%%%%%%%
%%%%%%%%%%%%%%%%%%%%%%%%%%%%%%%%%%%%%%%%%%%%%%%%
Moreover, we have:

\begin{proposition}
The axiom of time-orientation holds if and only
if $\tau^+_1\tau^-_2 \neq \tau^+_2\tau^-_1 $.

In that case
$$ \beta = \frac{\tau_2^- + \tau_2^+}{\tau^+_1\tau^-_2
   - \tau^+_2\tau^-_1} U^\dagger [D,U] -
     \frac{\tau_1^- + \tau_1^+}{\tau^+_1\tau^-_2 - \tau^+_2\tau^-_1}
     V^\dagger [D,V]
$$
\end{proposition}

\begin{proof}
A simple computation shows that
$ [D,U] V   =  {\lambda V[D,U]} $, and likewise $ [D,U] U  =  U [D,U]$.
Similarly for $[D,V]$ .\\
Accordingly, any one-form $\omega$ can be written as
$\omega = a_U [D,U] + a_V[D,V]$ with
$a_U,a_V \in \CA$. We can therefore make the Ansatz
$\beta =a_U [D,U] + a_V[D,V]$. Now, since $\beta$ commutes
with $\delta_1,\delta_2$ we immediately infer that $a_U = w U^\dagger$
and $a_V= z V^\dagger$ with $w,z \in \mathbb{R}$. (Reality follows
from $\beta^\dagger = -\beta$.) \\
The resulting linear equation for $w,z$ then has the above
solution, respectively it has no solution
if $\tau^+_1\tau^-_2 = \tau^+_2\tau^-_1 $.
\end{proof}

\begin{remark}
In might be very instructive to compute the metrics on the set of pure
states over $\CA$ corresponding to the above four parameter family of
Dirac-Operators. However, the space of pure states is explicitly
known only in the commutative case, $\lambda=1$, in which it is just
the two-dimensional Torus $T^2$. $U,V$ can then be identified with
$U=e^{i\varphi_1}, V=e^{i\varphi_2}$ for the usual ``coordinates''
$\varphi_1,\varphi_2 \in [0,2\pi]$.

We therefore consider only this commutative case in the
following. It is then possible to compute the metric as
follows: We may write
$$ D= \Gamma_1\delta_1 + \Gamma_2\delta_2, \qquad\qquad\qquad \Gamma_i =
   \left( \begin{array}{cc} 0 & \tau_i^+ \\ \tau_i^- & 0 \end{array}
   \right), \quad i= 1,2 .
$$
With this definition the components $g_{ij}$ of the metric with respect to the
coordinates $\varphi_1,\varphi_2$  are then given by the standard formula
$$ \Gamma_i\Gamma_j + \Gamma_j\Gamma_i= -2 g_{ij} \left( \begin{array}{cc} 1 & 0  \\  0 & 1 \end{array} \right).$$
(Note that $\delta_i = i\frac{\partial}{\partial\varphi_i}$ in this case.)
Here we obtain (combining the components $g_{ij}$ to a two-by-two matrix $g$),
$$ g = - \left( \begin{array}{cc} \tau_1^+\tau_1^- & \oh ( \tau_2^+\tau_1^-+ \tau_1^+\tau_2^-)  \\
               \oh ( \tau_2^+\tau_1^-+ \tau_1^+\tau_2^-)   &\tau_2^+\tau_2^-  \end{array} \right).$$
Thus, we immediately infer that (unless $\tau^+_1\tau^-_2 = \tau^+_2\tau^-_1$)
$$ {\rm det}(g) =\tau_1^+\tau_1^-\tau_2^+\tau_2^- -\frac{1}{4} ( \tau_2^+\tau_1^-+ \tau_1^+\tau_2^-)^2
   = - \frac{1}{4} ( \tau_2^+\tau_1^- - \tau_1^+\tau_2^-)^2 < 0. $$
Hence $g$ always has one positive and one negative eigenvalue and thus it
is always of Lorentzian signature.
\end{remark}

Let us now return to the case of generic $\lambda$.
\begin{proposition}
$[\langle D \rangle, [D,a]]$ is bounded for all $a\in\CA$.
\end{proposition}
We only sketch the proof:
\begin{proof}
First of all we observe that
\[ [\langle D \rangle, [D,U]] = \frac{1}{\sqrt{2}}
\left( \begin{array}{cc} 0 & \tau_1^+[\sqrt{\partial_+^2 + \partial_-^2},U] \\
\tau_1^- [\sqrt{\partial_+^2 +\partial_-^2},U]& 0 \end{array} \right) \]
and similarly for $V$. Due to the Leibniz rule for commutators it
is sufficient to consider the generators of $\CA$. The result then follows
by induction for all $a\in \CA$. \\
We therefore only need to prove the boundedness
of $[\sqrt{\partial_+^2 +\partial_-^2},U]$ and $[\sqrt{\partial_+^2 +\partial_-^2},V]$.
Since $\partial_\pm$ are derivations on the algebra, this is a well known fact, and
we need not repeat the somewhat lengthy proof here. Readers interested to see it
are referred to \cite{BFV}.
\end{proof}

\begin{proposition}
The axiom of orientation is fulfilled, i.e. $\gamma$ can be written as
a two-form which is the image of a Hochschild-cycle under $\pi$.
\end{proposition}
\begin{proof}
Note that  $\gamma = \beta \sigma$ where
\[ \sigma =  \left( \begin{array}{cc} 0 & 1 \\ 1& 0 \end{array} \right)
 = \frac{\tau_2^- - \tau_2^+}{\tau^+_1\tau^-_2 - \tau^+_2\tau^-_1} U^\dagger [D,U] +
   \frac{\tau_1^-  - \tau_1^+}{\tau^+_1\tau^-_2 - \tau^+_2\tau^-_1} V^\dagger [D,V]  . \]
Hence $\gamma$ is a two-form.
The computation that this two-form is the image of a Hochschild-cycle is lengthy but straightforward,
and completely analogous to that in \cite{BFV}. We therefor leave it to the reader.
\end{proof}

So we can finally claim
\begin{theorem}
The above data provide, if $ \tau^+_1\tau^-_2 \neq \tau^+_2\tau^-_1$,
irreducible ${\bf u(1)\oplus u(1)}$-equivariant Lorentzian spectral triples
over the noncommutative torus.
\end{theorem}

\begin{remark}
Similarly as in \cite{PaSiTo} we can describe the spectrum of
the Lorentzian Dirac operator on the noncommutative torus with
the choice of a different spin structure. We recall that the
four different spin structures were labelled by two numbers,
$\sigma_\pm$ which could take values $0$ or $\oh$. The above
presented case is for $\sigma_+=\sigma_-=0$, however, the
construction can be easily extended to the remaining situations.
The Dirac operator, for each of the spin structure, is:
$$ D \left(\begin{array}{ll}
0 & \tau_1^+ (n+ \sigma_+) + \tau_2^+ ( m + \sigma_-) \\
\tau_1^ - (n + \sigma_+) + \tau_2^- ( m + \sigma_-) & 0
\end{array} \right),
$$
with real parameters $\tau_1^\pm,\tau_2^\pm$.
\end{remark}

%%%%%%%%%%%%%%%%%%%%%%%%%%%%%%%%%%%%%%%%%%%%%%%%%%%%%%%%%%%%%%%%%%%
\section{The Lorentzian noncommutative isospectral 3-sphere}

In the three-dimensional, apart from the obvious case of
a three-dimensional torus we have another example of a compact
Lorentzian geometry: sphere. We look with $(1,2)$ signature,
so we look for the operators $\beta, J, D$ satisfying:
\begin{equation}
\begin{aligned}
\beta &= -\beta^\dagger, \;\;\; & \beta^2 & =-1, \\
J^2 &= 1, \;\;\; & J \beta &= \beta J, \\
D^\dagger &= \beta D \beta,\;\;\;& JD &= - DJ.
\end{aligned} \label{1plus2}
\end{equation}

\subsection{The isospectral deformation of the sphere}

Let $\lambda$ be a complex number of module one, which is not a
root of unity. The isospectral deformation of the three-sphere was
first presented in \cite{CoLa}, then in \cite{CoDV}. The general
result for the spectral triple construction of Drinfeld-type twists
as isospectral deformations and their equivariance was discussed
in \cite{Sitarz-eq}.

We use the description of the algebra of $S^3_\lambda$ as generated
by operators $a,b$ and their hermitian conjugates, which act on
the Hilbert space of square integrable functions on $S^3$. Using
the basis, we have the explicit formulae:

%%%%%%%%%%%%%%%%%%%%%%%%%%%%%%%%%%%%%%%%%%%%%%%%%%%%%%%%%%%%%
\begin{equation}
\begin{aligned}
\pi(a) \ket{l,m,n} &=& \lambda^{\oh(m-n)} \left(
\frac{\sqrt{l+1+m}\sqrt{l+n+1}}{\sqrt{2l+1}\sqrt{2l+2}}
\ket{l^+, m^+, n^-} \right. \\
&& \left. - \frac{\sqrt{l-m}\sqrt{l-n}}{\sqrt{2l}\sqrt{2l+1}}
\ket{l^-,m^+,n^-}\right) ,
\end{aligned}
\end{equation}
%%%%%%%%%%%%%%%%%%%%%%%%%%%%%%%%%%%%%%%%%%%%%%%%%%%%%%%%%%%%%
\begin{equation}
\begin{aligned}
\pi(b) \ket{l,m,n} &=& \lambda^{-\oh(m+n)} \left(
\frac{\sqrt{l+1+m}\sqrt{l-n+1}}{\sqrt{2l+1}\sqrt{2l+2}}
\ket{l^+, m^-, n^-} \right.\\
&& \left. + \frac{\sqrt{l-m}\sqrt{l+n}}{\sqrt{2l}\sqrt{2l+1}}
\ket{l^-,m^-,n^-}\right) ,
\end{aligned}
\end{equation}
%%%%%%%%%%%%%%%%%%%%%%%%%%%%%%%%%%%%%%%%%%%%%%%%%%%%%%%%%%%%%
where $l^\pm,m^\pm,n^\pm$ is a shortcut notation for
$l \pm \oh, m\pm\oh,n\pm\oh$.

The representation give above is equivariant with respect to
the Drinfeld twist of the $\CU(su(2)) \otimes \CU(su(2))$
Hopf algebra. We shall, however, consider the spinorial
representation of the algebra, which differs by the twisting
by a two-dimensional representation in the second $su(2)$.
So, in addition to doubling the Hilbert space, we need to take
into account that the second $u(1)$ action is different.
Therefore, the equivariant spinorial representation of
$S^3_\theta$ is diagonal but not just doubled:
\begin{equation}
 \pi(x) = \left( \begin{array}{cc} \pi_+(x) & 0
 \\ 0 & \pi_-(x) \end{array} \right),
\end{equation}
where $\pi_\pm$ differ from $\pi_0$ through the rescaling of
the generators:

$$
\begin{aligned}
\pi_\pm(a) &= \lambda^{\pm \frac{1}{4}} \pi_0(a), \\
\pi_\pm(b) &= \lambda^{\mp \frac{1}{4}} \pi_0(b),
\end{aligned}
$$

In our case, to get the Lorentzian spectral geometry we cannot
keep the entire symmetry as the isometry, and we need to reduce
it to a smaller one, which shall be the Drinfeld twist of
$\CU(u(1)) \otimes  \CU(su(2))$. This, however, shall be
used only when looking for the equivariant reality and the
Dirac operators.

The Krein-space structure operator $\beta$ commutes both with the
symmetries and with the representation of the algebra, hence it
must be diagonal:

\begin{equation}
\beta = \left( \begin{array}{cc} i & 0
\\ 0 & -i \end{array} \right).
\end{equation}

The reality structure $J$, which satisfies the relations
(\ref{1plus2}) must be off-diagonal:
\begin{equation}
J = \left( \begin{array}{cc} 0 & J_0^-
\\ J_0^+  & 0 \end{array} \right),
\end{equation}
where $J_0^\pm$ is a canonical equivariant antilinear map, which
maps the algebra to its commutant:
\begin{equation}
J_0^\pm \ket{l,m,n} = i^{2(m+n)} \ket{l,-m,-n}.
\end{equation}
Let us verify that $J^2=1$:
$$
\begin{aligned}
J^2 \ket{l,m,n,\pm}
&= J_0^\mp \left( i^{2(m+n)} \ket{l,-m,-n,\mp} \right) \\
&= i^{-4(m+n)} \ket{l,m,n,\pm} = \ket{l,m,n,\pm},
\end{aligned}
$$
where we have used that $m+n$ is always integer.

Next, we begin looking for all equivariant operators $D$, which
satisfy the order one condition. From equivariance with respect
to the full right symmetry and the left $u(1)$ part we
infer that the most general form of an equivariant operator
is:

\begin{equation}
\begin{aligned}
D \ket{l,m,n,+} &= d_{11}(l,m)\, \ket{l,m,n,+}
+ d_{21}(l,m) \ket{l,m+1,n,-}, \\
D \ket{l,m,n,-} &= d_{12}(l,m) \ket{l,m-1,n,+} + d_{22}(l,m) \ket{l,m,n,-}.
\end{aligned}
\end{equation}
{}From $JD=DJ$ condition we read:
\begin{equation}
\begin{aligned}
d_{11}(l,m)^* &= d_{22}(l,-m),\\
d_{21}(l,m)^* &= d_{12}(l,-m).
\end{aligned}
\end{equation}
For practical reasons it is more convenient to use the variables $l\pm m$,
so, instead of writing $d_{12}(l,m)$ we shall write $d_{12}(l+m,l-m)$.

Next, looking at $[JaJ,[D,b]]$, we first obtain:
\begin{equation}
\begin{aligned}
2 \sqrt{l-m+1} \, & d_{21}(l+m,l-m+1) =  \sqrt{l-m}\, d_{21}(l+m,l-m) \\
& + \sqrt{l-m+2} \, d_{21}(l+m,l-m+2),
\end{aligned}
\end{equation}
which has the solution:
$$ d_{21}(l+m,l-m) =  \frac{R(l+m)}{\sqrt{l-m}} + S(l+m)\sqrt{l-m}, $$
with arbitrary functions $R,S$.

Similarly, from one of the coefficient of $[JaJ, [D,b]]$ we read:
$$ 2 d_{11}(l-\oh,m+\oh)= d_{11}(l,m) + d_{11}(l-1,m+1), $$
which has the solution:
$$ d_{11}(l,m) = R'(l+m) + (l-m) \, S'(l+m), $$
with arbitrary functions $R',S'$.

Putting back these solutions into the formulae for the coefficients and
looking again at the order one condition we obtain relations:
$$ S'(l+m)-S'(l+1+m) = 0,$$
hence $S'(l+m)= S_0' = \hbox{const}$. and:
$$ S(l+m)\,\sqrt{l+2+m}=  S(l+1+m)\,\sqrt{l+1+m}, $$
which has the solution:
$$ S(l+m) = S_0 \sqrt{l+1+m}, $$
with the arbitrary multiplicative constant $S_0$.

Further, the condition for $R'(l+m)$ becomes:
$$ 2R'(l+m+1) = R'(l+m)+ R'(l+m+2),$$
which leads to
$$ R'(l+m) = R_1' + R_0' \, (l+m), $$
with $R_0',R_1'$ constant.

On the other hand, for $R$ we obtain:
$$ \sqrt{l+m}\, R(l+m-1) = \sqrt{l+m+1}\, R(l+m), $$
which has a solution:
$$ R(l+m) = \frac{R_0}{\sqrt{l+m+1}}. $$

Using this result and looking once again at the relations from the
order-one condition we obtain that $R'_0 = S'_0$ and $R_0$ must
vanish.

Hence the Dirac operator has the following form, when acting on
vectors $\ket{l,m,n,\pm}$:

\begin{equation}
\begin{aligned}
D \ket{l,m,n,+} &= i R m \,\ket{l,m,n,-} \\
& + S \sqrt{l+1+m} \sqrt{l-m} \, \ket{l,m+1,n,-}, \\
D \ket{l,m,n,-} &= - i R m\, \ket{l,m,n,+} \\
& + S^* \sqrt{l-m+1} \sqrt{l+m} \, \ket{l,m-1,n,+},
\end{aligned}
\label{Diracs3}
\end{equation}

where $S$ is arbitrary complex number and $R$ is real. It is easy
to verify that these restrictions arise from the conditions
$D^\dagger = \beta D \beta$ and $JD=-DJ$.

\begin{lemma}
The spectrum of the Dirac operator is:
$$ \lambda(l,m) =
-\frac{1}{2} i R \pm \frac{1}{2} \sqrt{|S|^2 (l + \oh)^2 - (|S|^2 + R^2)(m + \oh)^2}.
$$

We have the following possibilities:
\begin{itemize}
\item $R=0$: spectrum is real and symmetric:
$$ \lambda(l,m) = \pm |S| \sqrt{(l-m)(l+m+1)},$$
with $0$ being an eigenvalue with infinite multiplicity;

\item $S =0$: spectrum is pure imaginary:
$$ \lambda(l,m) = - \frac{1}{2} i R (1 \pm (1+2 m)),$$

\item $S \not=0$, $R \not=0$: spectrum might contains a pure
imaginary part and a complex part with imaginary part lying
on the $ -\oh i R$ axis.

The multiplicity of the eigenvalues depends on the ratio $\frac{|S|}{R}$,
if this is irrational then all eigenvalues have multiplicity $1$, whereas
in the rational case some of the eigenvalues might occur multiple
(but finite) number of times.
\end{itemize}
\end{lemma}

Direct calculations lead to the following result:

\begin{lemma}
The selfadjoint operator $\frac{1}{2} \langle D \rangle^2$ is has spectrum:
\begin{equation}
\hbox{Spec}(\langle D \rangle^2)
= \{ R^2 (m + \oh \pm \oh)^2 + |S|^2 (l-m) (l+m+1) \}
\end{equation}
and has compact resolvent if and only if $R|S| \not= 0$.
\end{lemma}
\begin{proof}
We restrict ourselves only to one set of eigenvalues. Taking
$m' = m+ \oh$ and $l' = l+\oh$ we can rewrite the formula:
$$ \lambda_{\langle D \rangle^2} = R^2 (m' + \oh)^2 +
|S|^2 (l'^2 - m '^2).$$
Then, it is clear that the number of such eigenvalues, which are less
then $\Lambda$ is finite. Indeed, both components must be less than
$\Lambda$ since $l'^2 \geq m'^2$, so:
$$ (m'+\oh) < \frac{\Lambda}{R}, \;\;\; l'^2 < \frac{\Lambda}{|S|} + m'^2,$$
so, unless $R=0$ or $|S|=0$ we get an estimate on $l$, and therefore for
each $\Lambda$ only finite number of eigenvalues are below it.
\end{proof}

\begin{lemma}
The following cycle:
$$ c = \sum_i c_i \ts c_i' =
\frac{i}{R} \left( a \ts A + b \ts B - A \ts a - B \ts b \right) $$
gives the time-orientability:
$$ \beta = \sum \pi(c_i) [D, \pi(c_i')]. $$
\end{lemma}

We can summarize our result:

\begin{theorem}
There exists a one-parameter family of $su(2) \otimes_\theta u(1)$
equivariant Lorentzian spectral geometries on the noncommutative
three sphere $S^3_\theta$, given by the Dirac operator \ref{Diracs3}.
\end{theorem}

\begin{remark}
It is worth noting that taking the $\lambda=1$ limit one recovers the
construction (albeit in a different framework) and eigenvalues of
the Lorentzian Dirac operator on the sphere, presented in
\cite{Baum}.
\end{remark}

\subsection{The reconstruction of the metric}

Having the explicit form of the Dirac operator we might attempt
to calculate the metric components, to see whether the obtained
Lorentzian structure on the three-sphere has no singularities.
Clearly, for the same reason as in the case of noncommutative torus,
we are limited to the $\lambda=1$ example.

It is convenient to use the basis of the left-invariant (hermitian)
one-forms on the 3-sphere:
$$
\begin{aligned}
\omega^1 &= b^* da - a db^* + b da^* - a^* db, \\
\omega^2 &= i \left( a db - bda - a^* db^* + b^* da^* \right), \\
\omega^3 &= i \left( b db^* + a^* da \right),
\end{aligned}
$$
then the inverse of the metric, calculated as
$$ g^{ij} = \frac{1}{2} \left( \pi(\omega^i) \pi(\omega^j)
+\pi(\omega^j) \pi(\omega^i) \right), $$
with $\pi(x\, dy) = \pi(x) [D,\pi(y)]$, becomes:
$$
g_{ij} = \frac{1}{2} \left(
\begin{array}{ccc}
-|S|^2 & 0      & 0 \\
0     & -|S|^2  & 0 \\
0     & 0      & \frac{1}{4} R^2
\end{array} \right)
$$
and is exactly the constant Berger-type metric with signature $(1,2)$
(the Euclidean part has negative sign), which was the starting point
of Helga Baum's approach \cite{Baum}.

\section{Isospectral deformations}

In the previous sections we have derived all equivariant Lorentzian
geometries on two examples of noncommutative manifold: the noncommutative
torus and the noncommutative three-sphere. In both examples we have
obtained the {\em isospectral deformation}, that is the Dirac operator
came out as the classical one. This should not be surprising, as one can
easily generalize the theorem for {\em isospectral deformations}
of \cite{CoLa} to the Lorentzian case, and we can claim:

\begin{proposition}[compare \cite{CoLa} Theorem 6]
Let $M$ be Lorentzian spin manifold and
$(C^\infty(M),\CH,D,\gamma,J,\beta)$ be the ingredients of
the associated Lorentzian spectral triple. We assume
that the isometry group of $M$ has rank at least $2$.
Then $M$ admits a natural one-parameter isospectral
deformation $M_\theta$.
\end{proposition}

\begin{proof}
All steps of the construction from \cite{CoLa}, section 5, can be
repeated in the Lorentzian case. Since $U(1) \times U(1)$ is a
subgroup if the isometry group, $D$ and $\gamma$ (in the even
case) and the Krein-space structure operator $\beta$ commute
with the action of this group. Hence, if we take for the deformed
spectral triple the same Krein-space structure $\beta$, and the
same Dirac operator $D$, we retain all relations and properties
of the triple with the exception of the orientability and the
time-orientability axioms.
\end{proof}

\begin{remark}
The orientability and time-orientability cannot be automatically
extracted from the commutative spectral triple data. As an example
one can take exactly $S^3_\theta$: in the classical case ($\theta=0$)
the cochain giving the orientability axiom might be chosen as
$$ a \ts a^* + b \ts b^*, $$
however this particular choice does not give a suitable orientability
for the deformation $S^3_\theta$.
\end{remark}

\section{The Lorentzian quantum sphere $SU_q(2)$}

The $\theta$-deformations are not the only one existing deformations
of the $3$-sphere. In the Euclidean version the quantum deformation
of $SU_q(2)$ and the related spectral geometry \cite{Naiad} were one
of the first examples of genuine noncommutative spectral geometries
beyond the isospectral examples.

We shall study here the possibility of obtaining a Lorentzian-type
spectral geometry for the $SU_q(2)$, seen as $S^3_q$. Of course,
the total symmetry of the quantum space $\CU_q(su(2)) \otimes
\CU_q(su(2))$ cannot be preserved, however, we might have a reduced
$\CU(u(1)) \otimes \CU_q(su(2))$ equivariance.

We construct the spectral triple using the spinorial equivariant
representation used in \cite{Naiad}. Still, though the representation
of $SU_q(2)$ is $\CU_q(su(2)) \ts \CU_q(su(2))$ equivariant, the
Dirac operator shall be only $\CU_q(su(2)) \ts u(1))$ equivariant.

We briefly recall the fundamentals of the representation
used in \cite{Naiad}.

For $j = 0,\oh,1,\frac{3}{2},\dots$, with $\mu = -j,\dots,j$ and
$n = -j-\half,\dots,j+\half$, we compose the pair of spinors:
\begin{equation}
\kett{j\mu n} := \begin{pmatrix} \ket{j\mu n\up} \\[2\jot]
\ket{j\mu n\dn} \end{pmatrix},
\label{eq:kett-defn}
\end{equation}
with the convention that the lower component is zero when
$n = \pm (j+\half)$ or $j = 0$.
Furthermore, a matrix with scalar entries,
$$
A = \begin{pmatrix} A_{\up\up} & A_{\up\dn} \\
A_{\dn\up} & A_{\dn\dn} \end{pmatrix},
$$
is understood to act on $\kett{j\mu n}$ by the rule
\begin{align}
A \ket{j\mu n \up}
&= A_{\up\up} \ket{j\mu n\up} + A_{\dn\up} \ket{j\mu n\dn}, \nn
\\
A \ket{j\mu n \dn}
&= A_{\dn\dn} \ket{j\mu n\dn} + A_{\up\dn} \ket{j\mu n\up}. \label{actonarrows}
\end{align}

The representation $\pi' := \pi \ts \id$ of $\CA$ is given by
$$
\begin{aligned}
\pi'(a) \,\kett{j\mu n}
&= a^+_{j\mu n} \kett{j^+ \mu^+ n^+}
 + a^-_{j\mu n} \kett{j^- \mu^+ n^+},
\nn \\[\jot]
\pi'(b) \,\kett{j\mu n}
&= \tilde{b}^+_{j\mu n} \kett{j^+ \mu^+ n^-}
 + \tilde{b}^-_{j\mu n} \kett{j^- \mu^+ n^-},
\nn \\[\jot]
\pi'(a^*) \,\kett{j\mu n}
&= \tilde{a}^+_{j\mu n} \kett{j^+ \mu^- n^-}
 + \tilde{a}^-_{j\mu n} \kett{j^- \mu^- n^-},
 \label{eq:spin-repn} \\[\jot]
\pi'(b^*) \,\kett{j\mu n}
&= \tilde{b}^+_{j\mu n} \kett{j^+ \mu^- n^+}
 + \tilde{b}^-_{j\mu n} \kett{j^- \mu^- n^+},
\nn
\end{aligned}
$$

where $\tilde{a}^\pm_{j\mu n}$ and $\tilde{b}^\pm_{j\mu n}$ are, up
to phase factors depending only on~$j$, the following triangular
$2 \times 2$ matrices:
$$
\begin{aligned}
\tilde{a}^+_{j\mu n} &= q^{(\mu+n-\half)/2} [j + \mu + 1]^\half
\begin{pmatrix}
q^{-j-\half} \, \frac{[j+n+\frac{3}{2}]^{1/2}}{[2j+2]} & 0 \\[2\jot]
q^\half \,\frac{[j-n+\half]^{1/2}}{[2j+1]\,[2j+2]} &
q^{-j} \, \frac{[j+n+\half]^{1/2}}{[2j+1]}
\end{pmatrix},
\nn \\[2\jot]
\tilde{a}^-_{j\mu n} &= q^{(\mu+n-\half)/2} [j - \mu]^\half
\begin{pmatrix}
q^{j+1} \, \frac{[j-n+\half]^{1/2}}{[2j+1]} &
- q^\half \,\frac{[j+n+\half]^{1/2}}{[2j]\,[2j+1]} \\[2\jot]
0 & q^{j+\half} \, \frac{[j-n-\half]^{1/2}}{[2j]}
\end{pmatrix},
\nn \\[2\jot]
\tilde{b}^+_{j\mu n} &= q^{(\mu+n-\half)/2} [j + \mu + 1]^\half
\begin{pmatrix}
\frac{[j-n+\frac{3}{2}]^{1/2}}{[2j+2]} & 0 \\[2\jot]
- q^{-j-1} \,\frac{[j+n+\half]^{1/2}}{[2j+1]\,[2j+2]} &
q^{-\half} \, \frac{[j-n+\half]^{1/2}}{[2j+1]}
\end{pmatrix},
\label{eq:spin-coeff}
\\[2\jot]
\tilde{b}^-_{j\mu n} &= q^{(\mu+n-\half)/2} [j - \mu]^\half
\begin{pmatrix}
- q^{-\half} \, \frac{[j+n+\half]^{1/2}}{[2j+1]} &
- q^j \,\frac{[j-n+\half]^{1/2}}{[2j]\,[2j+1]} \\[2\jot]
0 & - \frac{[j+n-\half]^{1/2}}{[2j]}
\end{pmatrix},
\nn
\end{aligned}
$$
and the remaining matrices are the hermitian conjugates
$$
\tilde{a^*}^\pm_{j\mu n} = (\tilde{a}^\mp_{j^\pm \mu^- n^-})^\dagger,  \qquad
\tilde{b^*}^\pm_{j\mu n} = (\tilde{b}^\mp_{j^\pm \mu^- n^+})^\dagger.
$$

It is, however, convenient to use the approximate representation
from \cite{Naiad}, that is representation up to compact operators.

We have:

\begin{lemma}
The operator
$$
\beta \ket{j\mu n\up} =  i \ket{j\mu n\up}, \;\;\;
\beta \ket{j\mu n\dn} = -i \ket{j\mu n\dn},
 $$
commutes with the algebra up to compact operators and
satisfies the requirements for the Krein-structure:
$\beta^2=-i, \beta=-\beta^\dagger$.
\end{lemma}
and
\begin{lemma}
The following operator defined for $-j-\oh < n < j+ \oh$, $j>0$ as:
\begin{equation}
\begin{aligned}
D \ket{j,m,n,\up} &= (i r_\up j +i R_\up)  \,\ket{j,m,n,\up} \\
& + i S (j+n+\oh) q^{j-2n} \left( \frac{[j-n+\oh]}{[j+n+\oh]} \right)^\oh
\, \ket{j,m,n,\dn}, \\
D \ket{j,m,n,\dn} &= (-i r_\dn j - i R_\dn)  \,\ket{j,m,n,\dn} \\
&  -i S (j+n+\oh) q^{j-2n} \left( \frac{[j-n+\oh]}{[j+n+\oh]} \right)^\oh
 \, \ket{j,m,n,\up},
\end{aligned}
\label{Diracsuq2}
\end{equation}
where $R,r,S$ are real parameters, and on the remaining elements
of the basis:
\begin{equation}
D \ket{j,m, \pm(j+\oh),\up} = (r_\up j)  \,\ket{j,m,\pm(j+\oh),\up},
\end{equation}
is $\CU_q(su(2)) \ts u(1)$ invariant, $\beta$-self-adjoint and has
bounded commutators with the algebra elements.
\end{lemma}

\begin{proof}
Clearly the diagonal part of $D$ is the same (up to bounded corrections)
as in \cite{Naiad}, so we might restrict ourselves only to the
off-diagonal part, which we call $D_o$. From the equivariance,
$D_o$ must have the form:
$$ D_o  \ket{j,m,n,\up} = d_o(j,n)  \ket{j,m,n,\dn}, $$.

The commutator $[D, \pi(a)]$ reads:
$$
\begin{aligned}
[D,\pi(a)]  \ket{j,m,n,\up} &=
\left(  d_o(j^+,n^+)  \tilde{a}_{jmn}^+{\up\up} - d_o^*(j,n)
\tilde{a}_{jmn}^+{\dn\dn} \right)
 \ket{j^+,m^+,n^+,\dn} \\
& + d_o(j^+,n^+) \tilde{a}_{jmn}^+{\dn\up} \ket{j^+,m^+,n^+,\dn} \\
& + \left(  d_o(j^-,n^+)  \tilde{a}_{jmn}^-{\up\up} - d_o^*(j,n)
\tilde{a}_{jmn}^-{\dn\dn} \right)
 \ket{j^-,m^+,n^+,\dn} \\
& + d_o(j^-,n^+) \tilde{a}_{jmn}^-{\dn\up} \ket{j^+,m^+,n^+,\dn}.
\end{aligned}
$$
Using the approximate representation, we see that these relations
lead to the requirement that the following expressions remain
bounded:
$$ q^{j+m} q^{j+n} (d_o(j-\oh,n+\oh)-d_o(j,n)),$$
and
$$ \sqrt{1-q^{2j+2m+2}}
\left( \sqrt{1-q^{2j+2n+3}} d_o(j+\oh,n+\oh)-d_o(j,n) \sqrt{1-q^{2j+2n+2}} \right)
$$
First, we observe that if $d_o(j,n) = q^{j-n} c_o(j,n)$ and
$c_0(j,n) q^{j}$ is bounded then the first expression is certainly bounded
(of the order $q^{2j}$) and the second leads to the requirement that:
$$ c_o(j+\oh,n+\oh) - c_o(j,n)$$
is bounded.

This, however, is possible if at most:
$$ c_o(j,n) \sim  a_o (j+n) f(j-n) + g(j-n), $$
where $a_o$ is a constant and $f(j-n)$ is bounded. As for the function
$g$ we need only to guarantee that its growth is not too fast,
so that the other estimates are valid.

In particular we can choose $f \equiv 1, g \equiv 0$, so that
$d_o(j,n) = (j+n+\oh) q^{j-n}$ or $f \equiv 0, g(j-n) = j-n +\oh$,
however it is convenient to rewrite both expressions by perturbing
slightly functions by a bounded component,

In fact we can show that both:

$$
\begin{aligned}
&(j+n +\oh) q^{j-2n} \frac{\sqrt{[j-n+\oh]}}{\sqrt{[j+n+\oh]}} \\
&(j-n +\oh) q^{j} \frac{\sqrt{[j+n+\oh]}}{\sqrt{[j-n+\oh]}}
\end{aligned}
$$
satisfy the requirements.

Indeed, we have:

$$ \sqrt{\frac{[j-n+\oh]}{[j+n+\oh]}} = q^{n}
\frac{\sqrt{ 1 - q^{2j-2n+1}}}{\sqrt{ 1 - q^{2j+2n+1}}} $$

so that:

$$ (j+n +\oh) q^{j-2n} \frac{\sqrt{[j-n+\oh]}}{\sqrt{[j+n+\oh]}}
= q^{j-n} (j+n+\oh) \frac{\sqrt{ 1 - q^{2j-2n+1}}}{\sqrt{ 1 - q^{2j+2n+1}}} $$

and

$$ (j-n+\oh) q^{j} \frac{\sqrt{[j+n+\oh]}}{\sqrt{[j-n+\oh]}}
= q^{j-n} (j-n+\oh) \frac{\sqrt{ 1 - q^{2j+2n+1}}}{\sqrt{ 1 - q^{2j-2n+1}}}$$

where it is clear that $\sqrt{ 1 - q^{2j-2n+1}}$ and
$\sqrt{1 - q^{2j+2n+1}}$ are bounded functions and their
proportion is also bounded.

In both cases in the $q \to 1$ limit one recovers an $su(2) \times u(1)$
invariant first-order differential operator. Out of these two
possible terms only the first one is unbounded, the second has
the form $x q^{-x}$, which is a bounded function and therefore
could be considered rather as a small perturbation.
\end{proof}

For this reason, we shall rather concentrate our investigation
on the other contribution, thus taking as $D$ (\ref{Diracsuq2}).
Assuming (for simplicity) $r_\up = - r_\dn$ and
$R_\up=R_\dn + r = \frac{3}{2} r$ we calculate the spectrum of $D$:

\begin{lemma}
The spectrum of $D$ is:
\begin{eqnarray*} \lambda_D & = & \oh \pm
\sqrt{(- r^2 (2j + 1 )^2 + S^2 q^{2j-4n} (j+n+\oh)^2 \frac{[j-n+\oh]}{[j+n+\oh]}},
\end{eqnarray*}
for
$-j \leq m \leq j, -j-\oh < n < j+\oh, j = 0, \oh, \ldots,$
and
\begin{eqnarray*} \lambda_D' & = & ir (2j + \frac{3}{2}), \end{eqnarray*}
if $-j \leq m \leq j, n = \pm (j+\oh), j = 0, \oh, \ldots$.
\end{lemma}

We calculate spectrum of $\langle D \rangle^2$:
\begin{lemma}
The operator $\langle D \rangle^2$ has compact
resolvent, and its approximate spectrum is:

$$ \lambda_{\langle D \rangle^2} =
\oh r^2 (j + 1 \pm \oh)^2 + S^2 q^{2(j-n)} (j+n+\oh)^2 + o(q^j),$$

For this reason it is clear that $\langle D \rangle^2$ has a
compact resolvent, as the number of its eigenvalues smaller
that any
$N>0$ is always finite.
\end{lemma}

\begin{remark}
Note that $\beta$ commutes with the algebra only up to compact
operators. This, however, is to be expected in the geometries,
which arise from $q$-deformations. The classical ($q \to 1$)
limit yields a first-order differential operator, which gives
the metric of a correct signature $(1,2)$ provided that $|S|^2$
is bigger than $\frac{1}{4}R^2$, then, however, $\beta$ still
does not commute with the algebra.
\end{remark}

\section{Conclusions and Outlook}

In this paper we have shown that equivariance may be used
to construct explicit examples of Lorentzian spectral triples
for classical manifolds and their isospectral deformations.

In the q-deformed case, we have been able to use the equivariance
to construct an unbounded Fredholm module, with $D$ and elements
of $\CA$ having bounded commutator, but, so far, we fail to establish
the order-one condition (even up to compact operators). However, it seems
to us, that the problem to find a suitable {\em order-one} $D$ obeying
is related to the reduction of the ``isometry group'' from
$\CU_q(su(2))\ts\CU_q(su(2))$ as used in \cite{Naiad} to
$\CU_q(su(2))\ts u(1)$. This itself, is an intriguing problem and
we shall investigate it in details (in an Euclidean setup) in a
forthcoming paper.

Nevertheless, one of the important reasons for presenting the $SU_q(2)$ is
the existence of the fundamental equivariant symmetry $\beta$, which can
only be chosen to commute with the algebra up to compact operators and
has no apparent classical limit. The assumption that the commutation
relations could be relaxed appears a very natural choice, at least for
$q$-deformed geometries. Still, the orientability axioms -- full in the
Euclidean case and additionally time-orientability in the Lorentzian case
are a puzzle in this example. The nonexistence of the evident classical
limit in the presented construction might suggest that going out of the
Euclidean setup could open even more possibilities for geometries (even
in the rough meaning of unbounded Fredholm modules) than expected.

Of course, if one  wants to construct new examples of noncommutative
Lorentzian spectral triples which may serve as candidates for models
of spacetime, so that, in particular, the eigenvalues of $D$ have
infinite degeneracy, then one has to extend the application of
equivariance to locally compact quantum groups. This, and in
particular, the cases which have the $q$-deformed Lorentz and Poincar\'e
symmetries, or with the renowned $\kappa$-deformation of the Poincar\'e
group, remains an important challenge for future work.
%%%%%%%%%%%%%%%%%%%%%%%%%%%%%%%%%%%%%%%%%%%%%%%%%%%%%%%%%%%%%%%%%%%
%%%%%%%{\bf Acknowledgements:}
%%%%%%%%%%%%%%%%%%%%%%%%%%%%%%%%%%%%%%%%%%%%%%%%%%%%%%%%%%%%%%%%%%%

%%%%%%%%%%%%%%%%%%%%%%
\end{document}